\documentclass[12pt]{article}
\pdfoutput=1
\usepackage{latexsym, graphicx,cite} 
\usepackage{amsmath}
\usepackage{amssymb}
\usepackage[colorlinks=true, linkcolor=blue, bookmarks=true]{hyperref}

 \newcommand{\arXiv}[1]{\href{http://www.arXiv.org/abs/#1}{#1}}

\makeatletter
\renewcommand\section{\@startsection {section}{1}{\z@}%
                                   {-3.5ex \@plus -1ex \@minus -.2ex}
                                   {2.3ex \@plus.2ex}%
                                   {\normalfont\large\bfseries}}
\renewcommand\subsection{\@startsection{subsection}{2}{\z@}%
                                     {-3.25ex\@plus -1ex \@minus -.2ex}%
                                     {1.5ex \@plus .2ex}%
                                     {\normalfont\bfseries}}
\makeatother

\newcommand{\beq}{\begin{equation}}
\newcommand{\eeq}{\end{equation}}

\newcommand{\de}{\delta}
\newcommand{\del}{\partial}
\newcommand{\cF}{{\cal F}}
\newcommand{\cP}{{\cal P}}
\newcommand{\te}{{\theta}}

\renewcommand{\wp}{\varphi}

\begin{document}

\begin{titlepage}
\begin{flushright}
\phantom{arXiv:yymm.nnnn}
\end{flushright}
\vfill
\begin{center}
{\Large\bf Hidden symmetries of the Higgs oscillator\vspace{2mm}\\ and the conformal algebra}    \\
\vskip 15mm
{\large Oleg Evnin$^{a,b}$ and Rongvoram Nivesvivat$^a$}
\vskip 7mm
{\em $^a$ Department of Physics, Faculty of Science, Chulalongkorn University,\\
Thanon Phayathai, Pathumwan, Bangkok 10330, Thailand}
\vskip 3mm
{\em $^b$ Theoretische Natuurkunde, Vrije Universiteit Brussel and\\
The International Solvay Institutes\\ Pleinlaan 2, B-1050 Brussels, Belgium}

\vskip 3mm
{\small\noindent  {\tt oleg.evnin@gmail.com, rongvoramnivesvivat@gmail.com}}
\vskip 10mm
\end{center}
\vfill

\begin{center}
{\bf ABSTRACT}\vspace{3mm}
\end{center}

We give a solution to the long-standing problem of constructing the generators of hidden symmetries of the quantum Higgs oscillator, a particle on a $d$-sphere moving in a central potential varying as the inverse cosine-squared of the polar angle. This superintegrable system is known to possess a rich algebraic structure, including a hidden $SU(d)$ symmetry that can be deduced from classical conserved quantities and degeneracies of the quantum spectrum. The quantum generators of this $SU(d)$ have not been constructed thus far, except at $d=2$, and naive quantization of classical conserved quantities leads to deformed Lie algebras with quadratic terms in the commutation relations. The nonlocal generators we obtain here satisfy the standard $su(d)$ Lie algebra, and their construction relies on a recently discovered realization of the conformal algebra, which contains a complete set of raising and lowering operators for the Higgs oscillator. This operator structure has emerged from a relation between the Higgs oscillator Schr\"odinger equation and the Klein-Gordon equation in Anti-de Sitter spacetime. From such a point-of-view, constructing the hidden symmetry generators reduces to manipulations within the abstract conformal algebra $so(d,2)$.

\vfill

\end{titlepage}


\section{Introduction}

In this article, we shall be concerned with hidden symmetries of a quantum particle on a $d$-sphere moving in a central potential varying as the inverse cosine-squared of the polar angle. This system was brought up and solved by Lakshmanan and Eswaran in \cite{LE} as a quantum-mechanical (0+1)-dimensional analog of the pion nonlinear sigma model (a field theory describing low-energy interactions of pions and based on a nonlinear realization of the chiral $SU(2)\times SU(2)$ symmetry). Rewriting this model in more conventional variables revealed that it is a quantum particle on a 3-sphere moving in the potential we mentioned above. Importantly, the explicit solution for the spectrum displayed in \cite{LE} indicated a very high energy level degeneracy, which could not be explained solely by the obvious rotational symmetries of the problem.

The system we are considering is often called the Higgs oscillator, because much of the interest in the algebraic structures arising from its quantization has been spurred by Peter Higgs in \cite{Higgs}, where an investigation into the causes of the high degeneracy of the energy levels displayed in \cite{LE} revealed a hidden $SU(d)$ symmetry. The follow-up paper \cite{Leemon} provided some further considerations at $d>2$. The presence of the $SU(d)$ symmetry was demonstrated at the classical level, and it was explained how the degenerate quantum energy levels fill multiplets of $SU(d)$. Attempts to construct the quantum generators of this hidden $SU(d)$ have met much less success, except for the solution at $d=2$ given in \cite{Higgs}. The reason is that the classical conserved quantities depend in a complicated way on the canonical variables and are not straightforwardly quantized due to ordering ambiguities. In the quantum context, large sets of operators commuting with the Hamiltonian can be constructed, but their commutation relations do not close to form a Lie algebra. Rather, one obtains Lie-like algebras whose commutation relations are deformed by quadratic terms. The problem of constructing quantum $SU(d)$ generators for the Higgs oscillator at $d>2$, to the best of our knowledge, has remained unsolved \cite{Leemon, interbasis}.

Much of the subsequent extensive literature has focused on rich algebraic structures arising from commutation relations of conserved quantities of the quantum Higgs oscillator, in particular, in relation to its superintegrability. A selection of papers (including those dealing with other quantum systems closely related to the Higgs oscillator) can be found in \cite{quadr, quadr1,quadr2,quadr3,quadr4,quadr5,quadr5a,quadr5b,quadr6,quadr7,quadr8,quadr9,BP}. A key theme in these articles is explorations of nonlinear deformations of Lie algebras arising in connection to quantization of nonlinear mechanical systems, with some connections \cite{quadr2,quadr5a} to quantum groups. Some very intriguing relations have been exposed in \cite{HNY,CL} between multicenter modifications of the Higgs oscillator (in which the potential is a sum of the Higgs oscillator potentials sourced by a set of specially arranged points on a sphere) and the Calogero model. (We would also like to mention further developments along similar lines \cite{CHLN} that appeared in preprint form concurrently with the present article.)

Our main purpose here will be to return to the unresolved question, originally formulated in \cite{Higgs,Leemon}, of constructing quantum generators of the Higgs oscillator's $SU(d)$ symmetries. The reason we can say something new in this regard is that a surprising connection has been revealed in \cite{EK,EN} between the Higgs oscillator Schr\"odinger equation and the Klein-Gordon equation in Anti-de Sitter spacetime (a maximally symmetric spacetime of constant negative curvature). The considerations in \cite{EK,EN} were predominantly motivated by the nonlinear stability questions in Anti-de Sitter spacetime \cite{BR,MRlectures,BizonReview,CEproc} and in particular the associated problem of selection rules \cite{CEV1,CEV2,Yang}. The geometrization of the Higgs oscillator provided by these considerations suggested a number of algebraic structures, explored in \cite{EN}, which would be difficult to guess from a purely quantum-mechanical perspective. The geometrical symmetry (isometry) group of the Anti-de Sitter  spacetime with $d$ spatial dimensions is $SO(d,2)$, which also happens to be the group of conformal transformations of the $d$-dimensional flat Minkowski spacetime. With this in mind, it is hardly surprising that the corresponding $so(d,2)$ conformal algebra of operators acting on the Higgs oscillator states can be constructed. In fact, a part of this algebra contains a complete set of raising and lowering operators that allows one to generate any energy eigenstate starting from the vacuum \cite{EN}.

In what follows, we shall review the basic formulation of the quantum Higgs oscillator and develop its conformal algebra structure (section 2), describe how this conformal algebra can be used to construct the $SU(d)$ symmetry generators (section 3), and finally solve an auxiliary problem necessary to make our constructions completely explicit and involving a family of polynomials associated to the conformal algebra (section 4). Even though the algebraic structures we utilize are fundamentally motivated by the connections to the Anti-de Sitter geometry, we have chosen to make the bulk of our presentation completely self-contained from a purely quantum-mechanical perspective and avoid any essential references to the higher-dimensional geometrization that underlies our formalism. All the operator structures we need are introduced through their explicit action on the Higgs oscillator states, and their properties are verified by direct computations. Readers interested in connections to the Anti-de Sitter geometry are requested to consult \cite{EN}.


\section{The Higgs oscillator and its conformal algebra}

The Higgs oscillator is a particle on a $d$-sphere moving in a potential $V(\te)\sim 1/\cos^2\te$, with $\te$ being the polar angle. It is a spherical analog of the isotropic harmonic oscillator in flat space \cite{Higgs}, a maximally superintegrable system (i.e., a system admitting $2d-1$ globally defined functionally independent integrals of motion) all of whose classical orbits are closed.

The quantum Higgs oscillator is described by the stationary Schr\"odinger equation
\beq
\left(-\Delta+\frac{(2\de-d)^2-1}{4\cos^2\te}\right)\Psi=E\Psi,
\label{schrod}
\eeq
where $\Delta$ is the $d$-sphere Laplacian. The normalization of the potential is given in terms of the arbitrary parameter $\de$ satisfying $\de\ge d/2$, and the choice of this parametrization is motivated by our forthcoming construction of the conformal algebra, in which $\de$ plays a simple role. The motion is confined to the hemisphere $\te<\pi/2$ since the potential blows up on the equator. Because of the spherical symmetry (rotations around the point $\te=0$) the variables completely separate in spherical coordinates $(\te,\Omega)$, with $\Omega$ collectively denoting the remaining $(d-1)$ coordinate angles, and a complete solution for the spectrum can be obtained \cite{LE,Higgs,Leemon}:
\begin{align}
&\Psi_{Nlk}=(\cos\te)^{\de-\frac{d-1}2}\,(\sin\te)^l\, P^{\left(l+\frac{d}{2}-1,\de-\frac{d}{2}\right)}_{(N-l)/2}(\cos 2\te)Y_{lk}(\Omega),\label{Psi}\\
&E_{N}=(N+\de)^2-\frac{(d-1)^2}{4}\label{energ}.
\end{align}
The energy levels depend only on $N$, but not on the angular momentum labels $l$ and $k$. This creates a very large level degeneracy which is precisely what had prompted the investigations of \cite{Higgs,Leemon} and the identification of the hidden $SU(d)$ symmetry. With respect to this symmetry, energy level number $N$ fills a fully symmetric rank $N$ tensor multiplet. For each value of $N$, $l$ runs over all integers of the same parity as $N$ with $l\le N$ (such that $(N-l)/2$ is a positive integer). The label $k$ collectively denotes all the azimuthal quantum numbers distinguishing the states within a given angular momentum multiplet. $Y_{lk}$ are the standard spherical harmonics on a $(d-1)$-dimensional sphere. $P_{(N-l)/2}$ are the Jacobi polynomials with the measure parameters indicated by the superscripts. They are polynomials of degree $(N-l)/2$ whose detailed form will not be important in our considerations below. (When comparing the above expressions to, say, \cite{Higgs,Leemon}, one has to keep in mind that a number of conventions differ. Thus our $N$ is called $n$ in \cite{Higgs,Leemon}, while our $d$ is called $N$, the energy normalization differs by a factor of 2, and we are working with a $1/\cos^2\te$ potential, whereas a $\tan^2\te$ potential differing by a constant energy shift is employed in \cite{Higgs,Leemon}.)

While spherical coordinates are convenient for explicitly solving the differential equation (\ref{schrod}), other coordinate systems are much more suitable for displaying the algebraic structures of the problem that will be of interest for us here. The so-called `gnomonic' coordinates have been used in \cite{Higgs}, as they make it easy to write explicitly a large set of classical conserved quantities corresponding to (\ref{schrod}). The same coordinates also emerged naturally in \cite{EN} while examining connections of (\ref{schrod}) to the Klein-Gordon equation in Anti-de Sitter spacetime. Gnomonic coordinates correspond to labelling points on a hemisphere by their images on a flat hyperplane tangent at $\te=0$ under the projection from the center of the sphere. Mathematically, these coordinates are defined as $x^i=n^i(\Omega)\tan^2\te$, where $n^i(\Omega)$ is the unit vector pointing in the direction $\Omega$.

For our present purposes, another coordinate system will be even more useful. This coordinate system is given by an orthogonal projection of the hemisphere $\te<\pi/2$ on the flat equatorial hyperplane $\te=\pi/2$. If a unit sphere is embedded in a flat $(d+1)$-dimensional space as $(\xi^0)^2+\xi^i\xi^i=1$, the orthogonal coordinates are simply given by $\xi^i$, whereas in terms of the spherical coordinates $\xi^i=n^i(\Omega)\sin^2\te$. The sphere metric is 
\beq
ds^2=\left(\de_{ij}+\frac{\xi^i\xi^j}{1-\xi^2}\right)d\xi^id\xi^j.
\label{orthmetric}
\eeq
The $d$ coordinates $\xi^i$ vary over the unit ball $\xi^2<1$. This coordinate system actually takes us back to the very beginning of the Higgs oscillator story, as it was employed in \cite{LE} to set up the problem, though the solutions were only presented in terms of separation of variables, as in (\ref{Psi}), which hides a good deal of the underlying structure.

The sphere Laplacian in the orthogonal coordinates (\ref{orthmetric}) is $\Delta=\del_i\del_i-\xi^i\xi^k\del_i\del_k-d\xi^i\del_i$, and the Schr\"odinger equation (\ref{schrod}) takes the form
\beq
\left(-\del_i\del_i+\xi^i\xi^k\del_i\del_k+d\xi^i\del_i+\frac{(2\de-d)^2-1}{4(1-\xi^2)}\right)\Psi=E\Psi.
\label{orthschrod}
\eeq
Every eigenstate at the energy level number $N$ above the ground state for this equation is of the form
\beq
\Psi_N(\xi)=(1-\xi^2)^{\frac{2\de-d+1}{4}}\,\cP_N(\xi),
\label{PsiP}
\eeq
where $\cP_N$ is a polynomial of degree $N$ in $\xi^i$ satisfying the equation
\beq
\left(-\del^2+(\xi^i\del_i)^2+2\de \xi^i\del_i\right)\cP_N=(N^2+2\de N)\cP_N.
\label{eqP}
\eeq
While (\ref{eqP}) follows directly from (\ref{orthschrod}), the form of the eigenstates (\ref{PsiP}) can be verified, say, by taking the complete basis given by (\ref{Psi}) and rewriting it in terms of orthogonal coordinates. (Note that $\cos2\te=1-2\xi^i\xi^i$, while $\sin^l\te\, Y_{lk}(\Omega)= (\xi^i\xi^i)^{l/2}Y_{lk}(\Omega)$ are homogeneous polynomials of degree $l$ in $\xi^i$ known as the `solid harmonics.')
$\cP_N$ can be thought of as a family of orthogonal polynomials in a $d$-dimensional unit ball with respect to the measure $(1-\xi^2)^{\de-d/2}d\xi$. (This is a direct consequence of $\Psi_N$ satisfying orthogonality relations with respect to the standard invariant measure $(1-\xi^2)^{-1/2}d\xi$.) Such polynomials are analyzed in detail in \cite{DunklXu}. Explicit solutions (\ref{Psi}) allow one in principle to construct a complete basis for $\cP_N$ in terms of the standard Jacobi and Legendre polynomials, though this will not be of direct interest for us here. We remark that the notation $\cP_N$ does not represent a unique polynomial, but rather any such polynomial emerging from a Higgs oscillator energy eigenstate at the energy level number $N$ via (\ref{PsiP}). Construction of solutions to (\ref{eqP}) starts from specifying the coefficients of the monomials of degree $N$ contained in $\cP_N$, which are completely arbitrary. After that, (\ref{eqP}) fixes the coefficients of all the lower degree monomials.

The structure displayed by the equations (\ref{orthschrod}-\ref{eqP}) in orthogonal coordinates facilitates the introduction of a conformal algebra acting on the Higgs oscillator eigenstates. (This construction originally emerged in \cite{EN} from rather different reasoning, but we shall present it here in a minimal, directly verifiable form independent from the higher-dimensional geometrical considerations of \cite{EN}.)

First, we introduce what shall become the `dilatation generator' $D$, defined through its simple action on the energy eigenstates given by
\beq
D\Psi_N=(N+\de)\Psi_N.
\label{defD}
\eeq
In particular, the meaning of our parameter $\de$ is revealed as the eigenvalue of the dilatation generator on the vacuum state $|0\rangle=(1-\xi^2)^{\frac{2\de-d+1}{4}}$. Since the energy corresponding to the level number $N$ is a quadratic function of $N$ given by (\ref{energ}), $D$ can be expressed through the Higgs oscillator Hamiltonian $H=-\del_i\del_i+\xi^i\xi^k\del_i\del_k+d\xi^i\del_i+\{(2\de-d)^2-1\}/\{4(1-\xi^2)\}$ as
\beq
D=\sqrt{H+\frac{(d-1)^2}{4}}.
\label{defDsqrt}
\eeq
The fact that a square root is involved in the above formula for $D$ may make one worried whether this operator exists. Heuristically, one should feel safe because under the relation between the Higgs oscillator and the Klein-Gordon equation in AdS spacetime \cite{EK,EN}, $D$ becomes an ordinary time translation in AdS, which is a perfectly well-defined first order differential operator. (Similar remarks apply to other conformal algebra constituents we define below.) At a technical level, we give in appendix \ref{appD} an explicit formula for the action of $D$ in terms of well-defined solutions to the Klein-Gordon equation in AdS.

We further introduce the raising operators 
\beq
P_k=i\xi^k \left(D+\frac{d-1}2\right)-i\left(\del_k-\xi^k\xi^l\del_l\right),
\label{Pdef}
\eeq 
and the lowering operators 
\beq
M_k=i\xi^k \left(-D+\frac{d-1}2\right)-i\left(\del_k-\xi^k\xi^l\del_l\right).
\label{Mdef}
\eeq
(These formulas are simply orthogonal coordinate expressions for the operators $L^{\mbox{\scriptsize(Higgs)}}_{+k}$ and $L^{\mbox{\scriptsize(Higgs)}}_{-k}$ given in terms of gnomonic coordinates at the end of \cite{EN}. We will however verify their relevant properties directly in our considerations below, in a way that should make consulting \cite{EN} unnecessary for readers that are interested specifically in the Higgs oscillator, and not in its connections to the Anti-de Sitter spacetime.) One can furthermore give the action of (\ref{Pdef}-\ref{Mdef}) on the energy eigenstates (\ref{PsiP}):
\begin{align}
&P_k\Psi_N=-i(1-\xi^2)^{\frac{2\de-d+1}{4}}\left(\del_k-\xi^k\xi^l\del_l -(N+2\de)\xi^k\right)\cP_N,\label{Psiup}\\
&M_k\Psi_N=-i(1-\xi^2)^{\frac{2\de-d+1}{4}}\left(\del_k-\xi^k\xi^l\del_l +N\xi^k\right)\cP_N.\label{Psidown}
\end{align}
The states (\ref{Psiup}-\ref{Psidown}) are themselves manifestly of the form (\ref{PsiP}) with $\cP_N$ replaced by another polynomial. In (\ref{Psiup}), this new polynomial is of degree $N+1$. In (\ref{Psidown}), it may appear that the new polynomial is of degree $N+1$, but there is in fact a cancellation turning it into a polynomial of degree $N-1$. Namely, $\del_k \cP_N$ is of degree $N-1$, whereas the remaining two terms individually increase the degree of each monomial in $\cP_N$ by 1. However the combination $\xi^k(\xi^l\del_l -N)$ gives zero when acting on any monomial of degree $N$ in $\cP_N$. There are furthermore no monomials of degree $N-1$ in $\cP_N$ because all terms in $\cP_N$ have the same parity under the reflection $\xi\to-\xi$. (One simple way to see it is to notice that all states at the same energy level in (\ref{Psi}-\ref{energ}) have the same reflection parity $(-1)^l$, since $l$ can only vary in steps of 2 within the same energy level.) Therefore, the highest degree term that $\xi^k(\xi^l\del_l -N)$ can produce acting on $\cP_N$ is again of degree $N-1$. It can be verified by brute force algebra that the polynomials $\left(\del_k-\xi^k\xi^l\del_l -(N+2\de)\xi^k\right)\cP_N$ and $\left(\del_k-\xi^k\xi^l\del_l +N\xi^k\right)\cP_N$ solve (\ref{eqP}) with $N$ replaced by $N+1$ and $N-1$, respectively. The bottom line is that $P_k$ and $M_k$ are raising and lowering operators moving any energy eigenstate to another energy eigenstate one level higher or one level lower in the energy spectrum: $P_k\Psi_N=\tilde\Psi_{N+1}$, $M_k\Psi_N=\tilde\Psi_{N-1}$.

It turns out that the operators $D$, $P_k$ and $M_k$, supplemented with the angular momentum operator
\beq
L_{kl}=i\left(\xi_k\del_l-\xi_l\del_k\right)
\label{Ldef}
\eeq
form the conformal algebra $so(d,2)$:
\begin{align}
&[D,P_k]=P_k,\qquad [D,M_k]=-M_k,\qquad [P_k,M_l]=2(iL_{kl}-\de_{kl}D),\label{conf1}\\
&[D,L_{kl}]=[P_k,P_l]=[M_k,M_l]=0,\label{conf2}\\
&[P_k, L_{mn}]=i\left(\de_{mk}P_n-\de_{nk}P_m\right),\qquad [M_k, L_{mn}]=i\left(\de_{mk}M_n-\de_{nk}M_m\right),\label{conf3}\\
&[L_{kl},L_{mn}]=i\left(\de_{mk}L_{nl}-\de_{nk}L_{nm}+\de_{ml}L_{kn}-\de_{ln}L_{km}\right).\label{conf4}
\end{align}
The commutation relations of $D$ with $P$, $M$ and $L$ follow immediately from (\ref{defD}) and the fact that $P$, $M$ and $L$ change the energy level number by $+1$, $-1$ and $0$, respectively. The commutators $[P_k,P_l]$, $[M_k,M_l]$ and $[P_k,M_l]$ can be recovered by a brute force evaluation,\footnote{Care has to be exercised when using (\ref{Psiup}-\ref{Psidown}) to deduce the commutation relations, since the value of $N$ appearing on the right-hand side of (\ref{Psiup}-\ref{Psidown}) changes with each successive application of $P$ or $M$.} say, using (\ref{Psiup}-\ref{Psidown}). The last two lines are an immediate consequence of the tensorial properties of $P$ and $L$ with respect to rotations (and can, of course, also be verified directly). While the emergence of the conformal algebra (\ref{conf1}-\ref{conf4}) may appear abstruse in our present context, it becomes completely natural once the relation to the Anti-de Sitter spacetime \cite{EK,EN} has been revealed (and this is in fact how the structure we describe had been originally discovered). The conformal algebra simply generates the geometric symmetries of the Anti-de Sitter spacetime and its appearance is inevitable, rather than surprising, in this sense. We refer the interested reader to \cite{EK,EN} for further details.

The set of raising and lowering operators $P_k$ and $M_k$ is complete in the sense that, acting with different arrangements of raising operators on the vacuum $|0\rangle=(1-\xi^2)^{\frac{2\de-d+1}{4}}$, one can obtain a complete basis of the Higgs oscillator states. The energy level number $N$ is spanned by states of the form $P_{i_1}\cdots P_{i_N}|0\rangle$. The number of independent states of this form is precisely the number of totally symmetric tensors of rank $N$ (recall that different $P_k$ commute with each other). It is known since \cite{Higgs, Leemon} that the energy level number $N$ fills a totally symmetric rank $N$ tensor multiplet of the hidden $SU(d)$ symmetry. The dimensionality of these representations directly matches the number of independent states of the form $P_{i_1}\cdots P_{i_N}|0\rangle$.

The conformal algebra we have displayed is a spectrum-generating algebra, rather than a symmetry algebra of the Higgs oscillator. While $D$ and $L$ commute with the Higgs oscillator Hamiltonian, $P$ and $M$ do not. These raising and lowering operators will however form a crucial ingredient for the construction of the hidden symmetry generators that is our primary goal in this treatment.


\section{The hidden symmetry generators}

Classical conserved quantities for the Higgs oscillator whose Poisson brackets form an $su(d)$ Lie algebra are known since \cite{Higgs}. The quantization of these conserved quantities is, however, problematic, since they include non-polynomial dependences on canonical variables, and there is no obvious way to resolve the ordering ambiguities. In \cite{Higgs,Leemon}, a simple set of second order differential operators that commute with the quantum Hamiltonian was introduced. In our present language, these quantum conserved quantities can be equivalently written as
\beq
S_{mn}=P_mM_n+P_nM_m.
\eeq
Any product of one $P$ and one $M$ does not change the energy and therefore commutes with the Hamiltonian. The same can be said about any other product of equal numbers of $P$ and $M$. The problem with all these operators is not that they are not conserved (they are), but that their commutation relations among themselves do not form a Lie algebra. In order to close properly into an $SU(d)$ Lie algebra (after being supplemented with $L_{mn}$), $S_{mn}$ should possess commutators $[S_{mn},S_{kl}]$ linear in $L$. In fact, these commutators are deformed by quadratic terms involving products of $S$ and $L$. For the relatively simple case of $d=2$, it was explained in \cite{Higgs} how to modify $S$ in order to produce generators that satisfy the $su(2)$ Lie algebra. (We shall briefly revisit this material from our present angle in appendix \ref{app2d}.) For $d>2$, we are not aware of any solutions available in the literature. Our aim here is to demonstrate that the conformal algebra (\ref{conf1}-\ref{conf4}) allows one to construct symmetry generators satisfying the standard $su(d)$ Lie algebra for any $d$.

We start by observing the following auxiliary relation:
\beq
\sum_N \frac{1}{N!} P_{i_1}\cdots P_{i_N}|0\rangle\langle 0|M_{i_1}\cdots M_{i_N}=\cF(D,L^2),
\label{Fdef}
\eeq
where $\cF$ is some function\footnote{We thank Alexei Morozov for pointing out that $\cF$ is related to what has been studied in mathematical literature, starting with \cite{shapovalov}, under the name of the Shapovalov form. For our limited practical applications in this article, it will be easier to recover the concrete properties of $\cF$ we need by brute force analysis, rather than attempting to extract them from complicated general expressions.} of the dilatation generator $D$ and the square of the angular momentum $L^2\equiv L_{mn}L_{mn}/2$. That this relation holds can be verified by the following argument. Consider the action of the left-hand side of the above relation on a state at level $\tilde N$ given by $A_{i_1\cdots i_{\tilde N}} P_{i_1}\cdots P_{i_{\tilde N}}|0\rangle$. The result is $A'_{j_1\cdots j_{\tilde N}}P_{j_1}\cdots P_{j_{\tilde N}}|0\rangle$, where
\beq
A'_{j_1\cdots j_N}\equiv\frac{1}{N!} \langle 0|M_{j_1}\cdots M_{j_N}P_{i_1}\cdots P_{i_{N}}|0\rangle A_{i_1\cdots i_{N}}.
\label{Atransf}
\eeq
Any fully symmetric tensor $A_{i_1\cdots i_{ N}}$ can be decomposed into traceless parts irreducible under the action of the rotational $SO(d)$ denoted as $T^{(0)}_{i_1\cdots i_{ N}}$, $T^{(2)}_{i_1\cdots i_{ N-2}}$, $T^{(4)}_{i_1\cdots i_{ N-4}}$, etc:
\beq
A_{i_1\cdots i_{ N}}=T^{(0)}_{i_1\cdots i_{ N}}+\de_{\{i_1i_2}T^{(2)}_{i_3\cdots i_{ N}\}}+\de_{\{i_1i_2}\de_{i_3i_4}T^{(4)}_{i_5\cdots i_{ N}\}}+\cdots,
\label{Adecomp}
\eeq
where all the $T$-tensors are completely traceless. The curly bracket notation implies complete symmetrization with respect to all the indices enclosed, i.e., for any tensor $C_{j_1\cdots j_M}$,
\beq
C_{\{j_1\cdots j_M\}}\equiv\,\, (M!)^{-1} \hspace{-8mm}\sum_{\hspace{2mm}\mbox{\scriptsize Perm:}\{j_i\}\to\{k_i\}}\hspace{-6mm}C_{k_1\cdots k_M}.
\eeq
If only the term involving $T^{(N-l)}$ is present in the above sum, the corresponding state, given by $A_{i_1\cdots i_{ N}}P_{i_1}\cdots P_{i_N}|0\rangle$, is at the energy level number $N$ and carries the angular momentum $L^2=l(l+d-2)$. Spherical symmetry (of the theory and the vacuum state) furthermore guarantees that each term in (\ref{Adecomp}) is an eigenstate, with an eigenvalue that only depends on $N$ and $l$, of the transformation (\ref{Atransf}), irrespectively of the precise form of the (traceless, fully symmetric) tensor $T^{(N-l)}$. One can thus write:
\begin{align}
\frac{1}{N!} \langle 0|M_{j_1}\cdots M_{j_N}P_{i_1}\cdots P_{i_{N}}|0\rangle&\de_{\{i_1i_2}\cdots  \de_{i_{N-l-1}i_{N-l}}T^{(N-l)}_{i_{N-l+1}\cdots i_{ N}\}}\label{Fnl}\\
&=\cF(N+\de,l(l+d-2)) \,\de_{\{j_1j_2}\cdots  \de_{j_{N-l-1}j_{N-l}}T^{(N-l)}_{j_{N-l+1}\cdots j_{ N}\}}\nonumber,
\end{align}
which establishes (\ref{Fdef}) and defines $\cF$ more explicitly. ($N+\de$ and $l(l+d-2)$ are the eigenvalues of $D$ and $L^2$, respectively.) While spherical symmetry and the structure of its irreducible representations mandate (\ref{Fnl}), it is also possible to understand the emergence of this formula by purely algebraic reasoning, based on evaluating the scalar product on the left-hand side by successive applications of the commutation relations (\ref{conf1}-\ref{conf4}). We shall comment on this further at the beginning of the next section.

Multiplying (\ref{Fdef}) by $\cF^{-1}$, acting on the vector $P_{j_1}\cdots P_{j_{\tilde N}}|0\rangle$ and comparing the two sides, one deduces the complementary relation
\beq
\langle 0|M_{i_1}\cdots M_{i_N}\cF^{-1}P_{j_1}\cdots P_{j_{ N}}|0\rangle={N!}\, \de^{i_1}_{\{j_1}\cdots\de^{i_N}_{j_N\}}\equiv\hspace{-6mm}\sum_{\hspace{2mm}\mbox{\scriptsize Perm:}\{j_i\}\to\{k_i\}}\hspace{-6mm}\de_{i_1k_1}\cdots \de_{i_Nk_N},
\eeq
where complete symmetrization is  applied to the $j$-indices (and is automatically enforced for the $i$-indices as well). Note that, being a function of $D$ and $L^2$, $\cF$ obviously commutes with $D$ and $L_{ij}$. 

Armed with the above algebraic preliminaries, one proceeds to construct the hidden $SU(d)$ generators in the form
\beq
N_{rs}=\cF^{-1/2}\left(P_r\cF M_s+P_s\cF M_r\right) \cF^{-1/2}.
\label{Ndef}
\eeq
These operators manifestly preserve the energy level, and hence commute with the Higgs oscillator Hamiltonian and generate symmetries. Rather than retracing the convoluted guesswork that had led us to the above expression, we shall simply check below by brute force that it satisfies the desired commutation relations. (To gain some intuition on how this formula works, it is useful to consider an analogous construction for the usual $d$-dimensional isotropic harmonic oscillator. In this case, $P_i$ and $M_i$ should be replaced by the standard creation-annihilation operators $a^\dagger_i$ and $a_i$, after which $\cF$ becomes equal to 1, and $N_{rs}$ turn into the usual symmetry generators $a^\dagger_r a_s+ a^\dagger_s a_r$, discussed in \cite{Higgs}, for instance, as an elementary analog of the Higgs oscillator problem. For the Higgs oscillator itself, $\cF$ is no longer 1, because the raising and lowering operators form a more complicated algebra. Nonetheless, the above expression for $N_{rs}$ still does the job.)

To verify that $N_{rs}$ (together with $L_{mn}$) form the standard $su(d)$ Lie algebra, we first consider
\beq
N_{pq}N_{rs}=\cF^{-1/2}\left(P_p\cF M_q\cF^{-1}P_r\cF M_s+(p\leftrightarrow q)+(r\leftrightarrow s)+{{p\leftrightarrow q}\choose{r\leftrightarrow s}}\right)\cF^{-1/2}.
\label{NN}
\eeq
The index substitution notation, e.g., $(p\leftrightarrow q)$ implies adding a term obtained from the first term ($P_p\cF M_q\cF^{-1}P_r\cF M_s$) by applying the specified index permutation.
For the first term inside the brackets, substituting the definition of $\cF$, one gets
\begin{align}
&P_p\cF M_q\cF^{-1} P_r\cF M_s\\
&=\sum_N (N!)^{-2}P_p P_{i_1}\cdots P_{i_N}|0\rangle\langle 0|M_{i_1}\cdots M_{i_N}M_q\cF^{-1} P_rP_{j_1}\cdots P_{j_N}|0\rangle\langle 0|M_{j_1}\cdots M_{j_N}M_s\nonumber\\
&=\sum_N \frac{(N+1)!}{(N!)^2}\de^{q}_{\{r}\de^{i_1}_{j_1}\cdots\de^{i_N}_{j_N\}}\,P_p P_{i_1}\cdots P_{i_N}|0\rangle\langle 0|M_{j_1}\cdots M_{j_N}M_s\nonumber\\
&=\sum_N \frac{1}{N!}\Big\{ \de_{rq} P_p P_{i_1}\cdots P_{i_N}|0\rangle\langle 0|M_{i_1}\cdots M_{i_N}M_s+{N}\,P_p P_r P_{i_1}\cdots P_{i_{N-1}}|0\rangle\langle 0|M_{i_1}\cdots M_{i_{N-1}}M_sM_q\Big\}.\nonumber
\end{align}
The second term in the last line cancels out in the commutators $[N_{pq},N_{rs}]$ that we are aiming to compute, i.e.,
\beq
P_p\cF M_q\cF^{-1} P_r\cF M_s-{{p\leftrightarrow r}\choose{q\leftrightarrow s}}=\sum_N \frac{1}{N!}\de_{rq} P_p P_{i_1}\cdots P_{i_N}|0\rangle\langle 0|M_{i_1}\cdots M_{i_N}M_s -{{p\leftrightarrow r}\choose{q\leftrightarrow s}}.
\eeq
Consider the action of the above expression on $\cF^{-1}P_{j_1}\cdots P_{j_N}|0\rangle$:
\begin{align}
&(P_p\cF M_q\cF^{-1} P_r\cF M_s-P_r\cF M_s\cF^{-1} P_p\cF M_q)\cF^{-1} P_{j_1}\cdots P_{j_N}|0\rangle\\
&\hspace{5mm}=\frac{1}{(N-1)!}\de_{rq} P_p P_{i_1}\cdots P_{i_{N-1}}|0\rangle\langle 0|M_{i_1}\cdots M_{i_{N-1}}M_s\cF^{-1} P_{j_1}\cdots P_{j_N}|0\rangle -{{p\leftrightarrow r}\choose{q\leftrightarrow s}}\nonumber\\
&\hspace{5mm}= \de_{rq} P_p(\de_{sj_1} P_{j_2}\cdots P_{j_N}+\de_{sj_2}P_{j_1}P_{j_3}\cdots P_{j_N}+\cdots)|0\rangle  -{{p\leftrightarrow r}\choose{q\leftrightarrow s}}\nonumber.
\end{align}
At the same time, from the standard commutation relations between $L_{ij}$ and $P_k$, one gets
\beq
L_{ps}P_{j_1}\cdots P_{j_N}|0\rangle=i(\de_{sj_1}P_pP_{j_2}\cdots P_{j_N}+\de_{sj_2}P_pP_{j_1}P_{j_3}\cdots P_{j_N}+\cdots)|0\rangle - (p\leftrightarrow s).
\label{Laction}
\eeq
Hence,
\begin{align}
&\Big(P_p\cF M_q\cF^{-1}P_r\cF M_s-P_r\cF M_s\cF^{-1} P_p\cF M_q\Big)+(p\leftrightarrow q)+(r\leftrightarrow s)+{{p\leftrightarrow q}\choose{r\leftrightarrow s}}\nonumber\\
&\hspace{3cm}=-i(\de_{qr}L_{ps}+\de_{ps}L_{qr}+\de_{rp}L_{qs}+\de_{sq}L_{pr})\cF.
\end{align}
Taking into account that $L_{ij}$ commutes with $\cF$, one can finally convert (\ref{NN}) to the desired commutation relation (note that our convention for the sign of $L$ is opposite to the one used in \cite{Higgs})
\beq
[N_{pq},N_{rs}]=-i(\de_{qr}L_{ps}+\de_{ps}L_{qr}+\de_{rp}L_{qs}+\de_{sq}L_{pr}).
\label{NNcomm}
\eeq
The commutation relations of $L_{rs}$ and $N_{pq}$ can be established by much simpler means. Since $L_{rs}$ commutes with $\cF$, one simply has to move $L$ past $P$ and $M$ in (\ref{Ndef}) using the standard conformal algebra, which yields
\beq
[N_{pq},L_{rs}]=i(\de_{rp}N_{sq}-\de_{sp}N_{rq}+\de_{rq}N_{ps}-\de_{sq}N_{pr}).
\label{NLcomm}
\eeq
The commutation relations (\ref{NNcomm}-\ref{NLcomm}), together with the commutation relations (\ref{conf4}) for the angular momentum $L_{rs}$, guarantee that $N_{pq}$ and $L_{rs}$ generate the standard $su(d)$ Lie algebra, cf. \cite{Higgs}. It follows that $N_{pq}$ provide the correct quantization of the classical conserved quantities of the Higgs oscillator  forming an $su(d)$ Lie algebra under taking Poisson brackets. Note that the trace $N_{pp}$ is not an independent generator, but a function of $D$ and $L^2$.


\section{Explicit form of the auxiliary function $\cF$}

Our construction of the hidden symmetry generators (\ref{Ndef}) is still not entirely explicit in the sense that it involves the operator $\cF(D,L^2)$ defined by an infinite sum in (\ref{Fdef}). We shall now give further consideration to (\ref{Fnl}) and recover an explicit form of the function $\cF$.

When evaluating the scalar product $\langle 0|M_{j_1}\cdots M_{j_N}P_{i_1}\cdots P_{i_{N}}|0\rangle$ on the left-hand side of (\ref{Fdef}) one simply commutes $M$ past $P$ using the conformal algebra (\ref{conf1}-\ref{conf4}). The commutation produces insertions of $D$ and $L$, all of which have to be commuted past $P$ as well, until they reach $|0\rangle$. Reaching $|0\rangle$, $M$ and $L$ produce 0, while $D$ produces $\de|0\rangle$. After all the $M$-operators have been moved to the right in this fashion, what remains is a tensor carrying the indices $i_1\cdots i_N,j_1\cdots j_N$, made entirely of the Kronecker $\de$-symbols with the coefficients being polynomials of degree $N$ in $\de$ (this is because $\de$ can only emerge from the action of $D$ on $|0\rangle$, and each pair of $M$ and $P$ produces at most one $D$ during the commutation process). The tensor is of course fully symmetric under permutations of the $i$-indices among themselves, and the $j$-indices among themselves.

Given that $\langle 0|M_{j_1}\cdots M_{j_N}P_{i_1}\cdots P_{i_{N}}|0\rangle$ is made entirely of the Kronecker symbols and fully symmetric, it is evident that any tensor of the form $\de_{\{i_1i_2}\cdots  \de_{i_{N-l-1}i_{N-l}}T^{(N-l)}_{i_{N-l+1}\cdots i_{ N}\}}$ used in (\ref{Fnl}) simply gets multiplied by a number when contracted with $\langle 0|M_{j_1}\cdots M_{j_N}P_{i_1}\cdots P_{i_{N}}|0\rangle$ (remember that $T$ is completely traceless). This is because any given Kronecker symbol in $\langle 0|M_{j_1}\cdots M_{j_N}P_{i_1}\cdots P_{i_{N}}|0\rangle$ cannot have both of its indices contracted with $T$ (which would give zero because of the tracelessness condition), while any other contraction returns $T$ itself, multiplied by a number (the ordering of the $j$-indices is completely unimportant because of the complete symmetrization). In the end, one obtains, as in (\ref{Fnl}), the original tensor multiplied by $\cF(N+\de,l(l+d-2))$. The latter is in the form of a degree $N$ polynomial in $\de$ that we choose to parametrize as follows:
\beq
\cF(N+\de,l(l+d-2))=2^N\wp_{Nl}(\de).
\eeq

We have evaluated the polynomials $\wp_{Nl}$ for $N\le 6$ symbolically using the FORM computer algebra system \cite{form}, obtaining the following results:
\begin{align}
&\wp_{00}(\de)=1,\qquad\wp_{11}(\de)=\de,\\
&\wp_{22}(\de)=\de(\de+1),\qquad\wp_{20}(\de)=\de(\de-{\textstyle\frac{d}2}+1),\\
&\wp_{33}(\de)=\de(\de+1)(\de+2),\qquad\wp_{31}(\de)=\de(\de+1)(\de-{\textstyle\frac{d}2}+1),\\
&\wp_{44}(\de)=\de(\de+1)(\de+2)(\de+3),\qquad
\wp_{42}(\de)=\de(\de+1)(\de+2)(\de-{\textstyle\frac{d}2}+1),\\
&\wp_{40}(\de)=\de(\de+1)(\de-{\textstyle\frac{d}2}+1)(\de-{\textstyle\frac{d}2}+2),\\
&\wp_{55}(\de)=\de(\de+1)(\de+2)(\de+3)(\de+4),\\ 
&\wp_{53}(\de)=\de(\de+1)(\de+2)(\de+3)(\de-{\textstyle\frac{d}2}+1)\\
&\wp_{51}(\de)=\de(\de+1)(\de+2)(\de-{\textstyle\frac{d}2}+1)(\de-{\textstyle\frac{d}2}+2),\\
&\wp_{66}(\de)=\de(\de+1)(\de+2)(\de+3)(\de+4)(\de+5),\\ 
&\wp_{64}(\de)=\de(\de+1)(\de+2)(\de+3)(\de+4)(\de-{\textstyle\frac{d}2}+1),\\
&\wp_{62}(\de)=\de(\de+1)(\de+2)(\de+3)(\de-{\textstyle\frac{d}2}+1)(\de-{\textstyle\frac{d}2}+2),\\
&\wp_{60}(\de)=\de(\de+1)(\de+2)(\de-{\textstyle\frac{d}2}+1)(\de-{\textstyle\frac{d}2}+2)(\de-{\textstyle\frac{d}2}+3).
\end{align}
The above expressions display a clear pattern which suggests the following general formula:
\beq
\wp_{Nl}(\de)=(\de)_{(N+l)/2}(\de-{\textstyle\frac{d}2}+1)_{(N-l)/2},
\label{pguess}
\eeq
where $(x)_n$ denotes the Pochhammer symbol, $(x)_n\equiv x(x+1)(x+2)\cdots(x+n-1)$.

We shall now construct an inductive analytic proof for our guess (\ref{pguess}), which trivially holds at $N=l=0$ (since $\langle 0|0\rangle=1$). The idea is to express the scalar product in (\ref{Fnl}) at level $N$ through a similar product at level $N-1$. Consider
\begin{align}
&M_jP_{i_1}\cdots P_{i_N}|0\rangle=2(\de_{ji_1}D-iL_{ji_1})P_{i_2}\cdots P_{i_N}|0\rangle +
2P_{i_1}(\de_{ji_2}D-iL_{ji_2})P_{i_3}\cdots P_{i_N}|0\rangle +\cdots\nonumber\\
&\hspace{1cm}=2(N+\de-1)(\de_{ji_1}P_{i_2}\cdots P_{i_N}+\de_{ji_2}P_{i_1}P_{i_3}\cdots P_{i_N}+\cdots)|0\rangle\nonumber\\
&\hspace{2cm}+2P_j(\de_{i_1i_2}P_{i_3}\cdots P_{i_N}+\de_{i_1i_3}P_{i_2}P_{i_4}\cdots P_{i_N}+\cdots\\
&\hspace{4cm}+\de_{i_1i_N}P_{i_2}\cdots P_{i_{N-2}}+\de_{i_2i_3}P_{i_1}P_{i_4}\cdots P_{i_N}+\cdots)|0\rangle\nonumber,
\end{align}
where we have used (\ref{defD}) and (\ref{Laction}). We then contract the $i$-indices with a tensor corresponding to a state of angular momentum $l$ at level $N$, which can be written as  $T_{\{i_1\cdots i_l}\de_{i_{l+1}i_{l+2}}\cdots\de_{i_{N-1}i_N\}}$, $T$ being fully symmetric and traceless. We obtain
\begin{align}
&  M_{j_1}P_{i_1}\cdots P_{i_{N}}|0\rangle T_{\{i_1\cdots i_l}\de_{i_{l+1}i_{l+2}}\cdots\de_{i_{N-1}i_N\}}=[2l(N+\de-1)T_{j_1i_2\cdots i_l}\de_{i_1i_{l+1}}\label{TNminus1}\\
&\hspace{3cm}+(N-l)(N-l+2\de-d)T_{i_1\cdots i_l}\de_{j_1i_{l+1}}]P_{i_1}\cdots P_{i_{l+1}}(P_kP_k)^{\frac{N-l}2-1}|0\rangle\nonumber.
\end{align}
One can now evaluate the left-hand side of (\ref{Fnl}) as a product of $\langle 0|M_{j_2}\cdots M_{j_N}$ and the above state, whose components are all at level $N-1$. Before doing so, we note that the tensor appearing inside the square brackets in (\ref{TNminus1}) is no longer traceless with respect to $i_1\cdots i_{l+1}$ and needs to be decomposed into traceless parts in order to deal with states of definite angular momentum. This decomposition can be attained by applying the following general formula (the index $j_1$ is not contracted with the $P$-operators and plays no role in this trace decomposition):
\beq
A_{\{i_1}T_{i_2\cdots i_M\}}=\left[A_{\{i_1}T_{i_2\cdots i_M\}}-{\textstyle\frac{M-1}{d+2(M-2)}}A_kT_{k\{i_1\cdots i_{M-2}}\de_{i_{M-1}i_{M}\}}\right]+\left[{\textstyle\frac{M-1}{d+2(M-2)}}A_kT_{k\{i_1\cdots i_{M-2}}\de_{i_{M-1}i_{M}\}}\right],
\eeq
where the combinations inside the square brackets are individually fully symmetric and traceless.\footnote{Systematic computations involving complicated symmetrizations are conveniently performed using diagrammatic techniques \cite{cvet}, though conventional methods suffice for our present limited purposes. Some very recent applications of such diagrammatic techniques and elaborate symmetrization operations to problems involving conformal symmetry can be found, e.g., in \cite{diagconf}.} We can now apply the same trace decomposition to (\ref{TNminus1}) to obtain
\begin{align}
&  M_{j_1}P_{i_1}\cdots P_{i_{N}}|0\rangle T_{\{i_1\cdots i_l}\de_{i_{l+1}i_{l+2}}\cdots\de_{i_{N-1}i_N\}}\label{Tseparate}\\
&\hspace{1mm}=\left[2l(N+\de-1)+{\textstyle\frac{l}{d+2l-2}}(N-l)(N-l+2\de-d)\right]T_{j_1i_1\cdots i_{l-1}}P_{i_1}\cdots P_{i_{l-1}}(P_kP_k)^{\frac{N-l}2}|0\rangle\nonumber\\
&+(N-l)(N-l+2\de-d)\left[T_{i_1\cdots i_l}\de_{j_1i_{l+1}}-{\textstyle\frac{l}{d+2l-2}}T_{j_1i_1\cdots i_{l-1}}\de_{i_li_{l+1}}\right]P_{i_1}\cdots P_{i_{l+1}}(P_kP_k)^{\frac{N-l}2-1}|0\rangle\nonumber.
\end{align}
The last line is a state of angular momentum $l+1$ at energy level $N-1$, whereas the line above it is a state of angular momentum $l-1$ at energy level $N-1$. If we assume that we know the values of $\cF$ at energy level $N-1$, (\ref{Fnl}) will immediately tell us how to evaluate the product of $\langle 0|M_{j_2}\cdots M_{j_N}$ with (\ref{Tseparate}), and hence the values of $\cF$ at energy level $N$. Specifically,
\begin{align}
& \frac{1}{2^{N-1}(N-1)!} \langle 0|M_{j_1}M_{j_2}\cdots M_{j_N}P_{i_1}\cdots P_{i_{N}}|0\rangle T_{\{i_1\cdots i_l}\de_{i_{l+1}i_{l+2}}\cdots\de_{i_{N-1}i_N\}}\nonumber\\
&\hspace{1cm}=\Big\{\wp_{N-1,l-1}\left[2l(N+\de-1)+{\textstyle\frac{l}{d+2l-2}}(N-l)(N-l+2\de-d)\right]\\
&\hspace{2cm}+\wp_{N-1,l+1}(N-l)(N-l+2\de-d)\left[1-{\textstyle\frac{l}{d+2l-2}}\right]\Big\}T_{\{i_1\cdots i_l}\de_{i_{l+1}i_{l+2}}\cdots\de_{i_{N-1}i_N\}}\nonumber.
\end{align}
Substituting our guess (\ref{pguess}) for the $\wp$-polynomials at level $N-1$ in the above expression, we discover that the right-hand side is simply $2N\wp_{Nl}T_{\{i_1\cdots i_l}\de_{i_{l+1}i_{l+2}}\cdots\de_{i_{N-1}i_N\}}$, therefore confirming that the guess still holds at level $N$. This completes our inductive proof that
\beq
\cF(N+\de,l(l+d-2))=2^N(\de)_{(N+l)/2}(\de-{\textstyle\frac{d}2}+1)_{(N-l)/2}.
\label{Ffinal}
\eeq


\section{Outlook}

We have developed a family of operators acting on the space of states  of the $d$-dimensional quantum Higgs oscillator, defined by (\ref{defD}), (\ref{Pdef}), (\ref{Mdef}), (\ref{Ldef}) and forming an $so(d,2)$ conformal algebra (\ref{conf1}-\ref{conf4}). Using this algebra, we have constructed explicit generators of the $SU(d)$ hidden symmetries, given by (\ref{Fdef}) and (\ref{Ndef}), commuting with the Higgs oscillator Hamiltonian and forming an $su(d)$ Lie algebra. We have then derived an explicit expression (\ref{Ffinal}) in terms of Pochhammer symbols for the function $\cF$ used to construct the operator $\cF(D,L^2)$ utilized in (\ref{Ndef}). Our solution to the hidden symmetry generator problem is thus complete. (We comment in appendix \ref{app2d} on the relation of this solution to the previously known construction specific to $d=2$.)

Viewed from the perspective of the conformal algebra, the generators (\ref{Ndef}) reside in its universal enveloping algebra. In this sense, what we have presented is a construction of a certain $su(d)$ substructure in the universal enveloping algebra of the abstract $so(d,2)$. It would be interesting to explore what implications our findings have for multicenter Higgs potentials \cite{HNY,CL}, which would connect us to the extensive algebraic studies of the Calogero models, and for the dynamics in the Anti-de Sitter spacetime, where the Higgs oscillator emerges \cite{EK,EN} from separation of variables in the Klein-Gordon equation. This latter direction includes questions related to the much-studied AdS/CFT correspondence.


\section{Acknowledgments}

We would like to thank Ben Craps, Puttarak Jai-akson, Chethan Krishnan and Joris Vanhoof for collaboration on related subjects. We have benefitted from discussions with Piotr Bizo\'n, Joaquim Gomis and Alessandro Torrielli.
The work of O.E. is funded under CUniverse
research promotion project by Chulalongkorn University (grant reference CUAASC). R.N. is supported by a scholarship under the Development and Promotion of Science and
Technology Talents Project (DPST) by the government of Thailand.


\appendix

\section{Dilatation generator and AdS evolution}\label{appD}

As we remarked in the main text, the fact that the dilatation generator $D$ is given as a square root of a differential operator by (\ref{defDsqrt}) may make one worried whether this operator is well-defined. This concern is simply resolved by the relation between the Higgs oscillator Schr\"odinger equation and the linear Klein-Gordon equation in anti-de Sitter spacetime (AdS) exposed in \cite{EK,EN}. We shall now demonstrate how to exploit this relation to define $D$ explicitly.

The Klein-Gordon equation on AdS is given by
\beq
\cos^{2}x\ \left(
-\partial_t^2\phi + \frac{1}{\tan^{d-1} x} \partial_x(\tan^{d-1} x \partial_x \phi)+\frac{1}{\sin^2x} \Delta_{\Omega_{d-1}}\phi \right)
-m^2 \phi  = 0,
\label{KGAdS}
\eeq
where $x$ varies between $0$ and $\pi/2$, $\Delta_{\Omega_{d-1}}$ is the standard Laplacian on a $(d-1)$-dimensional sphere, and $m$ is related to $\de$ by
\beq
\de=\frac{d}2+\sqrt{\frac{d^2}{4}+m^2}.
\eeq 
There is a one-to-one correspondence \cite{EK,EN} between the eigenmodes of this equation and Higgs oscillator energy eigenstates. In particular, if one constructs the solution of (\ref{KGAdS}) satisfying initial conditions
\beq
\phi(t=0,x=\te, \Omega)=\cos^{(d-1)/2}\te\,\Psi_N(\te,\Omega),\qquad \del_t \phi(t=0,x,\Omega)=0,
\eeq
where $\Psi_N$ is an energy eigenstate of the Higgs oscillator at energy level number $N$, then for all times
\beq
\phi(t,x=\te, \Omega)=\cos(\de+N)t\,\cos^{(d-1)/2}\te\,\Psi_N(\te,\Omega).
\eeq
Consequently, 
\beq
-\frac{\del_t\phi(t=2\pi)}{\sin(2\pi\de)\cos^{(d-1)/2}\te}
\eeq 
coincides with the desired action of $D$ on $\Psi_N$ given by (\ref{defD}). This can be used to define $D$ through solutions of the Cauchy problem for (\ref{KGAdS}). Namely, for a given $\Psi(\te,\Omega)$ we define $\phi^{(\Psi)}$ to be the solution of (\ref{KGAdS}) satisfying initial conditions
\beq
\phi^{(\Psi)}(t=0,x=\te, \Omega)=\cos^{(d-1)/2}\te\,\Psi(\te,\Omega),\qquad \del_t \phi(t=0,x,\Omega)=0.
\eeq
Subsequently, the action of $D$ on $\Psi$ is defined by
\beq
D\Psi=-\frac{\del_t\phi^{(\Psi)}(t=2\pi)}{\sin(2\pi\de)\cos^{(d-1)/2}\te}.
\eeq
Well-posedness of the Cauchy problem for the Klein-Gordon equation in AdS (or even asymptotically AdS) spacetimes is established in mathematical literature \cite{holzegel} and ensures that the above definition is valid (while it explicitly coincides with the action of $D$ on Higgs oscillator eigenstates given by (\ref{defD}) in the main text).

\section{Simplifications in two dimensions}\label{app2d}

The conformal algebra $so(d,2)$ simplifies considerably at $d=2$. This is because there is only one $L$-generator left, $L_{12}\equiv L$, and its commutation relations with $P$ and $M$ become very simple if one introduces $P_{\pm}=P_1\pm iP_2$ and $M_{\pm}=M_1\pm iM_2$, yielding
\beq
[L,P_{\pm}]=\pm P_{\pm},\qquad [L,M_{\pm}]=\pm M_{\pm}.
\eeq
In other words, $P_+$ and $P_-$ respectively raise and lower the angular momentum (while raising the energy level by 1) while $M_+$ and $M_-$ respectively raise and lower the angular momentum (while lowering the energy level by 1). Evidently, for any function $f$,
\beq
f(L)P_{\pm}=P_{\pm}f(L\pm1),\qquad f(L)M_{\pm}=M_{\pm}f(L\pm1).
\eeq
Similarly, for any function $f$,
\beq
f(D)P_{\pm}=P_{\pm}f(D+1),\qquad f(D)M_{\pm}=M_{\pm}f(D-1).
\eeq

We can now revisit the construction of the hidden symmetry generators (\ref{Ndef}). It is convenient to form two Hermitian conjugate combinations
\begin{align}
&N_{+}=\cF^{-1/2}(D,L^2) P_+\cF(D,L^2) M_+\cF^{-1/2}(D,L^2),\\
&N_{-}=\cF^{-1/2}(D,L^2) P_-\cF(D,L^2) M_-\cF^{-1/2}(D,L^2).
\end{align}
Since in the beginning one has three Hermitian components of $N$, namely $N_{11}$, $N_{22}$ and $N_{12}$, with $N_{11}+N_{22}$ being a function of $D$ and $L^2$ rather than an independent generator, $N_{+}$ and $N_{-}$ encode all the independent generators one needs. Finally, using the above commutation properties, one writes
\begin{align}
&N_{+}=P_+ M_+\big[\cF^{-1/2}(D,(L+2)^2)\cF(D-1,(L+1)^2)\cF^{-1/2}(D,L^2)\big],\\
&N_{-}=P_-M_-\big[\cF^{-1/2}(D,(L-2)^2)\cF(D-1,(L-1)^2) \cF^{-1/2}(D,L^2)\big].
\end{align}
$P_+M_+$ and $P_-M_-$ are second order differential operators commuting with the Higgs oscillator Hamiltonian. They correspond to $S_+$ and $S_-$ introduced in section 5.2 of \cite{Higgs}. The construction of the hidden symmetry generators $N$ from the second order differential operators $S$ by modifying the latter with functions of $D$ and $L^2$ (identically, functions of the Higgs oscillator Hamiltonian and $L^2$) is then directly parallel to the considerations of \cite{Higgs}. This method does not generalize straightforwardly to higher dimensions, where the procedure presented in the main text of this article must be employed.


\end{document}